\UseRawInputEncoding
\documentclass[aps,prl,twocolumn,longbibliography,amsmath,amssymb,superscriptaddress]{revtex4-1}
\usepackage{amsmath} 
\usepackage {units}
\usepackage{gensymb}
\usepackage{here}
\usepackage{graphicx} 
\usepackage{verbatim} 
\usepackage{color} 
\usepackage{bm}
\usepackage{subfigure} 
\usepackage{hyperref} 
\usepackage[normalem]{ulem} 
\usepackage{siunitx} 
\usepackage[pagewise]{lineno}

\begin{document}
\title{Controlling topological defects and contractile flow in confined nematic cell population}
\author{Ryo Ienaga}\affiliation{Department of Physics, Kyushu University, Motooka 744, Fukuoka 819-0395, Japan}
\author{Kazusa Beppu}
\affiliation{Department of Physics, Kyushu University, Motooka 744, Fukuoka 819-0395, Japan}\affiliation{Department of Applied Physics, Aalto University School of Science, Puumiehenkuja 2, Espoo, 02150, Finland}

\author{Yusuke T. Maeda}\email{ymaeda@phys.kyushu-u.ac.jp}
\affiliation{Department of Physics, Kyushu University, Motooka 744, Fukuoka 819-0395, Japan}
\date{\today}

\begin{abstract}
Topological defects in nematically aligned cell populations play a critical role in modulating collective motion, from microbial colonies to epithelial tissues. Despite the potential of manipulating such topological defects to control diverse self-organized structures and collective dynamics, defect manipulation in active matter remains an challenging area of research. In this study, we investigated the geometric control of defect positioning and alignment in a nematic cell population by imposing spatial constraints consisting of two or three overlapping circular boundaries. The confined cell population exhibited an ordered pairing of half-integer topological defects that remained stable even when the size of the spatial constraint was altered using geometric parameters. These defects also elicited robust contractile flow that induced a negative divergence in the velocity field of collective motion. Such net contractile flow can contribute to mechanical stimulation on confined cells, as evidenced by the stretched cell nucleus. Our geometry-based approach paves the way for controlling defect pairing, providing a deeper understanding of the interplay among geometry, topology, and collective dynamics.
\end{abstract}
\maketitle

\subsection{Introduction} The self-organized structure and dynamics of nonequilibrium systems are crucial areas of investigation in the physics of soft matter and biological systems. In recent years, the ordered collective motion of active matter, which moves by consuming chemical energy, has been thoroughly studied to understand the underlying physical principles and applied engineering approaches in complex biological systems \cite{ramaswamy, marchetti}. A defining characteristic of active matter is that, even if each individual active element is moving randomly, their orientation interactions can induce spontaneous symmetry breaking in the rotational direction at high densities, resulting in a group of elements exhibiting ordered collective motions \cite{vicsek}. Two types of active matter orientation interactions involve polar interactions, which align the front-back axes of motion, and nematic interactions, which are characterized by the absence of front-back asymmetry. Active nematic --- a class of active matter with nematic interactions \cite{amin1} --- exhibits a nematically organized orientation field in various biological systems such as dense bacterial colonies \cite{zhang1, yaman}, cytoskeletal and molecular motor complexes \cite{dogic2, bausch1}, epithelial cell sheets \cite{silberzan1, henkes2020, feng1}, nematic cell populations \cite{yeomans1, silberzan2, saw, kawaguchi, amin2}, and hydra morphogenesis \cite{keren}.

Nematic interacting particles or cells form an ordered orientation field in two-dimensional space; however, singularities with orientations cannot continuously change, leading to spontaneous symmetry breaking of the local nematic order. These elements, called topological defects, can be created, or annihilated to influence the structure and mechanics of the materials \cite{shankar2019, shankar2022}. In nematic liquid crystals, rod molecules are misaligned at defects with specific orientations and topological charges. However, what distinguishes active nematics from passive nematic liquid crystals is the spontaneous motility of defects coupled with ordered collective motion \cite{amin1, silberzan1, yeomans1, silberzan2}. The spontaneous motion of topological defects allows for collisions and pair formation, which can alter the local degree of orientation order, leading to either the growth of long-range orientation order or persistent driving of active turbulent motions. In addition, topological defects in nematic cell populations alter the local cell density through asymmetric frictional interactions that trap or deplete cells \cite{saw, kawaguchi}. Because topology-associated mechanics largely determine collective movements and local changes in cell density in the vicinity of topological defects, the understanding of active nematics can be extended to investigate the relationship between cell proliferation control and rheological properties in microbial colonies \cite{durham2021, shaewitz2021} and eukaryotic tissue epithelia \cite{saw, keren, li2022}. While our understanding of collective motion in the presence of defects has been significantly advanced, developing methods to control the defect position remains an understudied area in the investigation of the interplay between topology and dynamics in active nematics.

One promising approach to controlling the collective motion of active matter is through the manipulation of the cell population under a confined geometry. Previous studies on confined active matter have made progress in controlling the behavior of active matter through polar interactions. A dense bacterial suspension exhibits active turbulent motion in boundary-free conditions but enclosing this dense suspension in spatial confinements allows the extraction of the vortex rotation inherent in bacterial turbulence \cite{wioland1}. Additionally, by making ordered bacterial vortices interact with each other, controlling the pairing pattern of the rotating vortices in a geometrically designed manner is possible \cite{wioland2, beppu, beppu2, nishiguchi2}. This development has also led to the extension of the geometric control approach to active nematics; Growing evidence shows that the shape of the boundaries plays a key role in organizing the emergent structure of the active cytoskeleton \cite{beppu3, hardouin2022, shendruk2021}, cell monolayers \cite{silberzan4, lim1, segerer, silberzan3, yeomans4}, and myoblast cell morphogenesis \cite{roux1,roux2}. Furthermore, complex patterns of topological defects under free boundary conditions can be created by building external patterns in liquid crystal materials on the substrate \cite{guillamat2016, gardel2018, turiv2020}. However, the principles of the geometric control of topological defect pairing and how defect positioning alters collective motion via defect positioning remain an elusive area of study.

In this study, we aim to develop a physical strategy for controlling active nematics using a geometric design of spatial confinement. We present a microfabrication method that allows nematic cell populations of myoblast progenitor cells to move only within a spatially constrained boundary shape. This confinement can control both the positioning and heading angles of half-integer topological defects while preserving the net topological charge in the orientation field. Moreover, we observed that the ordered pairing of defects led to the emergence of stable contractile flow toward the geometric center of the confined cell population across a wide range of geometric parameters. The pair of defects facing each other produces an ordered collective motion in which cells flow out of the defects, creating a stable inward flow that exerts a mechanical stress on the group of cells. Therefore, the geometric control of defect pairing may provide new insights into the interplay of geometry, topology, and mechanics in nematic cell populations.

\subsection{Results}
\begin{figure*}[tb]
\begin{center}
\includegraphics[scale=0.67,bb=0 0 752 552]{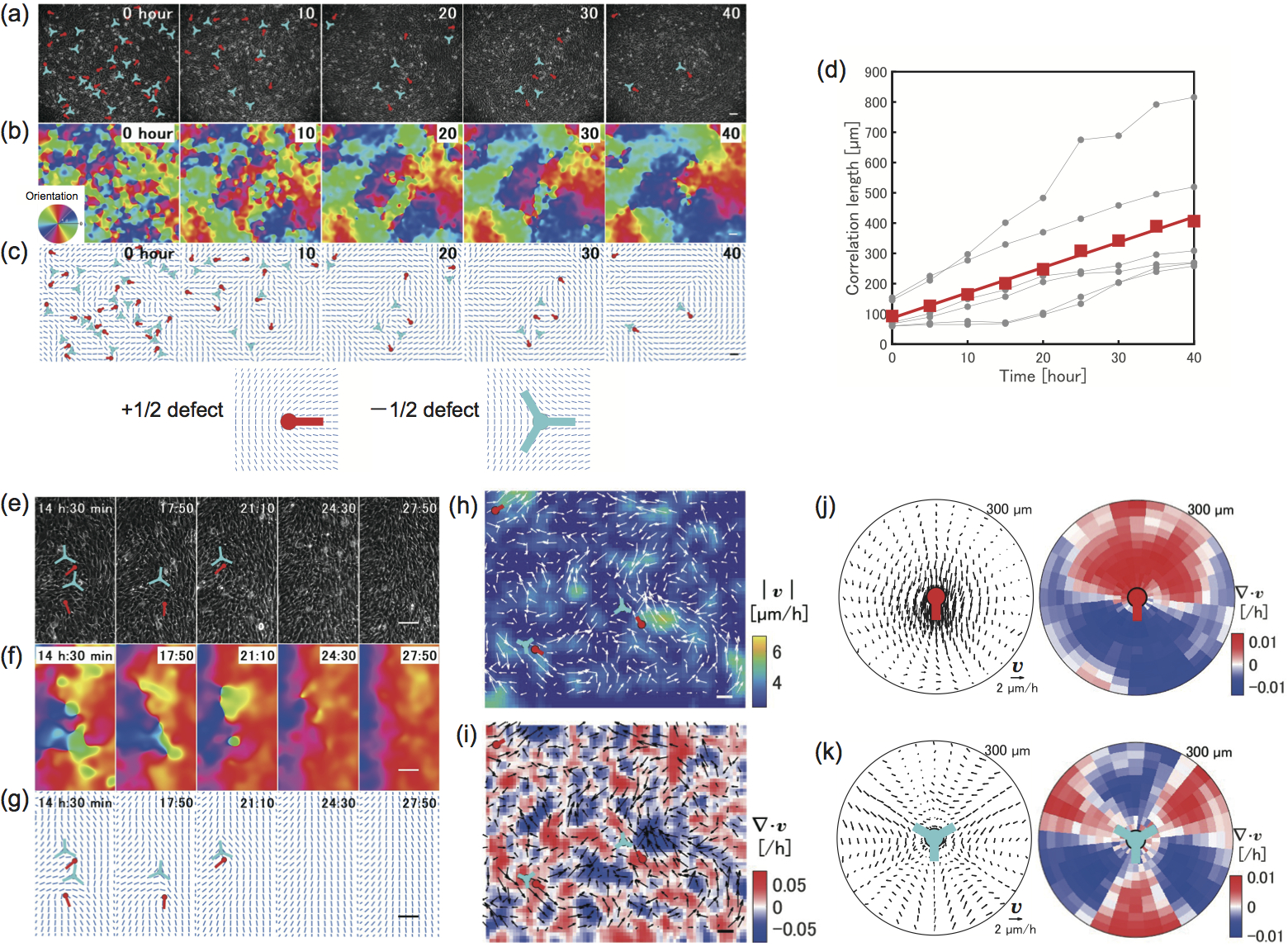}
\end{center}
\caption{\textbf{Collective motion and ordered orientation dynamics of myoblast cell  population.} Scale bars: \SI{100}{\micro\meter}.  \textbf{(a-c)} Collective motion and orientation dynamics of C2C12 cell population on a two-dimensional flat surface. \textbf{(a)} Phase contrast images. \textbf{(b)} The orientation field analyzed from (a). Color decodes the orientation field. \textbf{(c)} Corresponding director field analyzed from (a). The red symbol represents $+1/2$ topological defects, and the blue one represents $-1/2$ topological defects. \textbf{(d)} Time evolution of correlation length of orientation field, $\xi(t)$. Thin black lines represent the correlation length calculated from single experiments. Red dots represent the mean value and solid line is a fitting curve. \textbf{(e-g)} Annihilation of topological defects in C2C12 cell population. \textbf{(e)} Phase contrast images. \textbf{(f)} The orientation field analyzed from (e). \textbf{(g)} Corresponding director field analyzed from (e). \textbf{(h)} The speed $|\bm{v}|$ of collective motion of C2C12 cell population. The velocity field $\bm{v}$ is obtained from PIV analysis of phase contrast images. \textbf{(i)} The velocity divergence, $\bm{\nabla}\cdot\bm{v}$. \textbf{(j)} Spatial profiles of velocity $\bm{v}$ and the velocity divergence in the region from a $+1/2$ topological defect to \SI{300}{\micro\meter} radius. \textbf{(k)} Spatial profiles of velocity and the velocity divergence in the region from a $-1/2$ topological defect to \SI{300}{\micro\meter} radius.}\label{fig1}
\end{figure*}

First, we investigated the collective motion and associated orientation dynamics of a myoblast C2C12 cell population on a flat two-dimensional surface $\bm{r} = (r, \theta)$. The cultured cells had spindle-like cell morphology and, because of their polarized cell shape, could align their orientation heading angle through interaction with neighboring cells. Cell-cell adhesion was not strong in C2C12 cells, suggesting that their intercellular interactions were nematic in nature. The cells were grown in a culture dish, and their collective motion was recorded by time-lapse acquisition at 10-minute intervals for 48 h. The orientation field $\phi(\bm{r}, t)$ was calculated from phase-contrast microscope images. Initially, the cell population was relatively unorganized in their orientation, and singular points with local orientation gaps are distributed (FIG. \ref{fig1}(a), $t =$\SI{0}{\hour}). On a closed two-dimensional space, the topological charge of all defects must sum to $+1$, where a charge of $s$ indicates a defect that rotates the director field by $2\pi s$. Such singularities --- known as topological defects --- include the comet-shaped topological defect, in which the orientation phase deviates by $+\pi$（$s=+1/2$）, and another defect with three-fold symmetry, in which the orientation deviation is $-\pi$ ($s=-1/2$). These half-integer topological defects, comet-shaped $+1/2$ defects, and trifold $-1/2$ defects, were present in approximately equal numbers but seem to be distributed irregularly in space at earlier times (FIG. \ref{fig1}(a)-(c)).

The positions of these defects changed over time because of their spontaneous motion. As a result, the orientation of the cells gradually aligned, leading to the expansion of the orientation correlations (FIG. \ref{fig1}(b)-(c) and Movie S1). To investigate the growth of the nematic alignment order, we analyzed the time evolution of the correlation length $\xi(t)$ of the orientation field. The characteristic length scale of the autocorrelation function $C(dr, t) = \langle \cos2(\phi(r + dr, \theta, t) - \phi(r, \theta, t)) \rangle_{r, \theta}$ represents the correlation length $\xi(t)$ of the oriented field, and we defined $\xi(t)$ as the radial distance at which $C(dr,t)$ equaled $e^{-1}$. As shown in FIG. \ref{fig1}(d), the correlation length increased proportionally with time. Moreover, the spontaneous annihilation of $+1/2$ and $-1/2$ defects approaching each other occurred and then aligned with the cellular orientation (FIG. \ref{fig1}(e)-(g) and Movie S2). This is consistent with the region of the aligned orientation expanding over time (FIG. \ref{fig1}(d)), indicating that the defect pairing is key to organizing the ordered structures in nematic cell population.

\begin{figure*}[htb]
\begin{center}
\includegraphics[scale=0.88,bb=0 0 632 272]{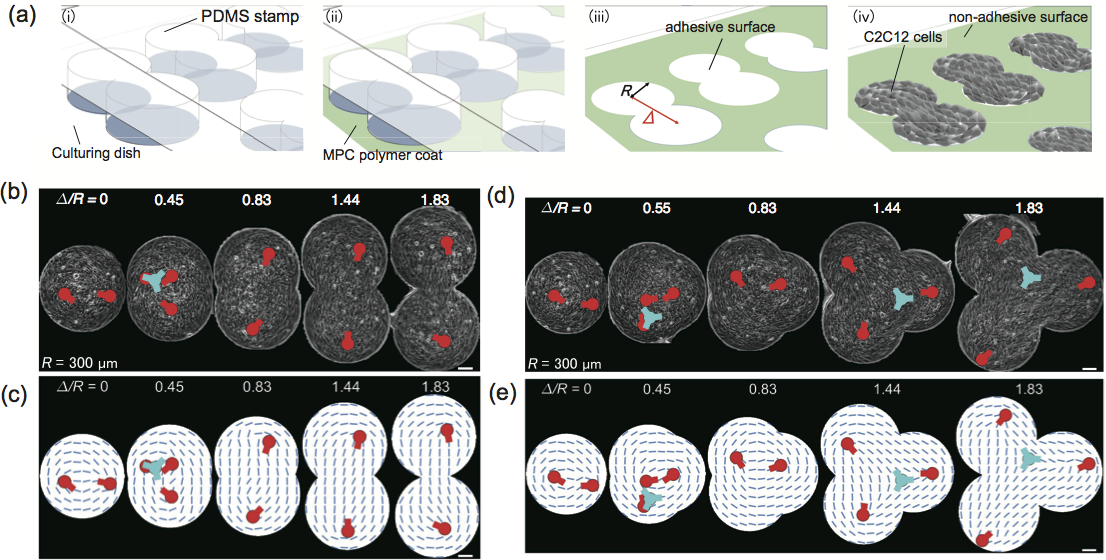}
\end{center}
\caption{\textbf{Patterning C2C12 cell population in confined overlapped circles and controlling topological defects pairing.} The red symbol represents $+1/2$ defect and the blue symbol $-1/2$ defect, respectively. Scale bars: \SI{100}{\micro\meter}. \textbf{(a)} Schematic illustration of microfabrication steps. Step (\rm{i}) Adheres the poly-dimethyl siloxane (PDMS) stamp weakly to the culture dish. Step (\rm{ii}) Coating of MPC polymer. Step (\rm{iii}) Make the designed adhesive region by removing the PDMS stamp. Step (\rm{iv}) Spread the C2C12 cells on the designed pattern and grow inside the adhesive region. \textbf{(b and c)} Phase contrast images of confined C2C12 cells at $t=\SI{48}{\hour}$. The confined C2C12 cells in (b) overlapped doublet circle pattern and (c) triplet circle pattern. \textbf{(d and e)} The director fields of confined C2C12 cells analyzed from (b) and (c). The local director patterns denote the localization of topological defects in confined geometries.The geometric constraint was a circle having radius $R=$\SI{300}{\micro\meter} with $\Delta/R$ = 0, 0.45, 0.83, 1.44, and 1.83.}\label{fig2}
\end{figure*}

The interplay between topological defects and collective motion distinguishes active nematics from passive liquid crystals. Reportedly, active forces associated with the orientation field drive collective motion around a topological defect \cite{amin1, shankar2019, shankar2022}. To clarify the coupling between the topological structure and velocity field of collective motion, we calculated the velocity field of collective motion $\bm{v}(\bm{x})$ by using particle image velocimetry (PIV). We found that the speed of the collective motion $|\bm{v}|$ is higher in the vicinity of $+1/2$ defects (FIG. \ref{fig1}(h)), but no clear indication of a global pattern existed in the velocity field. However, we found that the flow was generated from the head portion of the $+1/2$ defect in the comet shape to the tail portion, forming a characteristic flow field around the defect. To further analyze this defect-mediated flow, we calculated the velocity divergence $\nabla\cdot\bm{v}$ in a radius of $R=\SI{300}{\micro\meter}$ region centered on a single isolated $+1/2$ defect. The sign of the divergence reversed after the $+1/2$ defect because collective motion occurred in the direction of flow from the head side of this comet-shaped defect and in the direction of flow out on the tail side (FIG. \ref{fig1}(i)). Furthermore, the $-1/2$ defect showed three-fold symmetric velocity divergence, consistent with the fact that the orientation around the defect has three-fold symmetry (FIG. \ref{fig1}(k)), indicating a significant coupling between collective motion and the topological structure in a dense cell population.

\begin{figure*}[htb]
\begin{center}
\includegraphics[scale=0.81,bb=0 0 632 272]{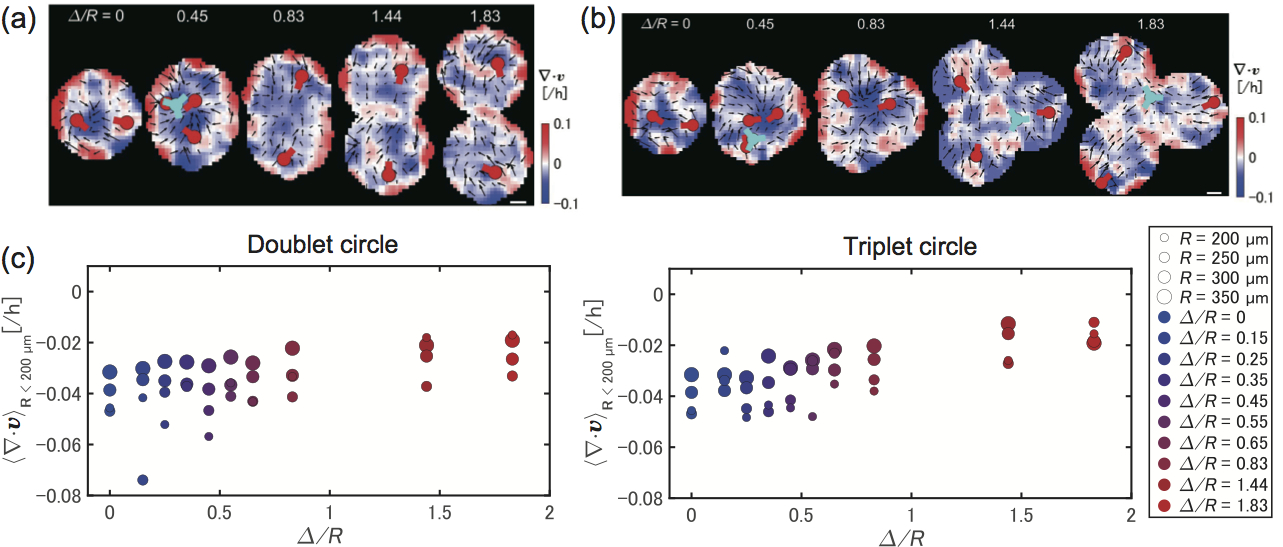}
\end{center}
\caption{\textbf{Negative velocity divergence, contractile flow, and topological defects pairing in confined C2C12 cells.} Scale bars: \SI{100}{\micro\meter}. \textbf{(a and b)} Velocity divergence $\bm{\nabla}\cdot\bm{v}$ in confined C2C12 cells. The cell population is confined in (a) doublet circle boundary and (b) triplet circle boundary with the geometric parameters $\Delta/R$ = 0, 0.45, 0.83, 1.44, and 1.83. \textbf{(c)} The velocity divergence $\bm{\nabla}\cdot\bm{v}$ in confined C2C12 cells in doublet circle boundary (left) and triplet circle boundary (right). We calculated the average velocity divergence within a radius of \SI{200}{\micro\meter} from the center of each circle. The geometric constraint was a circle having radius $R=$200, 250, 300 and \SI{350}{\micro\meter}, and we measured ten patterns with $\Delta/R=$0, 0.15, 0.25, 0.35, 0.45, 0.55, 0.65, 0.83, 1.44, and 1.83.}\label{fig3}
\end{figure*}

To further study the mechanism by which long-range orientation order emerges in C2C12 cell populations, we designed a geometrically controlled pairing of topological defects on a flat surface using the reverse micro-contact printing method (FIG. \ref{fig2}(a)). Geometric confinement consisted of a doublet or triplet circle geometry, in which two or three circles with the same radius $R$ overlapped at a distance $\Delta$ (FIG. \ref{fig2}(a)). The collision angle $2\Psi$ of the cells moving along the boundary can be geometrically defined by the non-dimensional parameter $\Delta/R = 2\cos\Psi$ \cite{beppu, beppu2, beppu3}. In active nematics, the positions of defects inside the constrained region determines the pattern of collective motion. In particular, the net charge of the defects inside the closed region is limited to $+1$. Therefore, we examined the geometry dependence of the defect positioning and the resulting collective motion on $\Delta/R$.

By analyzing phase-contrast images of a confined cell population, we found that the cellular orientation exhibited a bipolar-like orientation pattern in which the cells aligned in the long-axis direction of the doublet circle (FIG. \ref{fig2}(b) and Movie S3). As the distance between the centers of the circles, $\Delta$, increased at a constant radius, $R$, a single $+1/2$ topological defect was located within one circular compartment. The two $+1/2$ defects were aligned in a pattern, with their tails facing each other along the long axis of the doublet circle space (FIG. \ref{fig2}(c)), indicating that a nematically-ordered pattern was realized under this geometrical shape. Similarly, in the triplet circular boundary, in which three circles overlap, the cells showed three-fold symmetry and a trifurcated branching orientational pattern connecting the centers of the circles (FIG. \ref{fig2}(d) and Movie S4). Three-fold symmetry appeared as $\Delta$ increased (FIG. \ref{fig2}(e)): The three $+1/2$ defects were placed with their tails facing each other on the inside, while the $-1/2$ defect was placed near the geometric center of the constraint region, forming the basis of three-fold symmetry.

\begin{figure*}[htb]
\begin{center}
\includegraphics[scale=0.48,bb=0 0 1042 402]{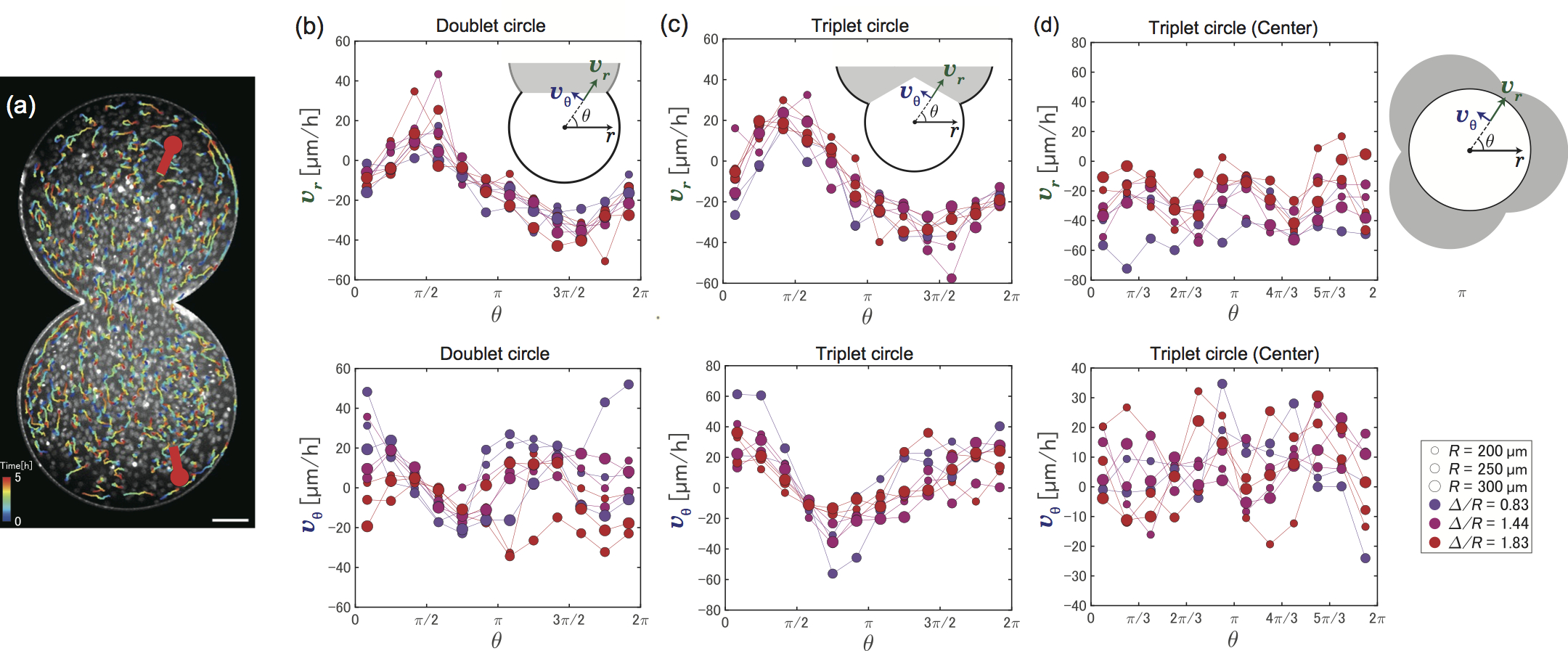}
\end{center}
\caption{\textbf{Contractile flow and angular dependence of collective motion}. \textbf{(a)} The trajectories of tracked cell nuclei in confined C2C12 cells at the geometry of $\Delta/R=1.83$, visualized using SiR-DNA. The color change of the trajectory represents the evolution over time. Scale bar: \SI{100}{\micro\meter}. \textbf{(b-d)} The radial velocity ($\bm{v_r}$, top) and angular velocity ($\bm{v_{\theta}}$, bottom) of cells are analyzed from nucleus tracking and are averaged over a fan-shaped region of interest divided every 25 degree. In (b), the velocity profiles in the doublet circle boundary are shown, with the center of one of the two overlapping circles being the origin of the polar coordinate $(r, \theta)$. The angle $\theta$ is set so that the center of the other circle is on the extension line of $\theta = \pi/2$, and the positive sign of $\bm{v_r}$ indicates that the cell nucleus moves away from the center of the circle. In (c), the velocity profiles in the triplet circle boundary are shown, with the center of one of the three overlapping circles being the origin of the polar coordinate. $\theta$ is set so that the geometric center of the three overlapping circles is on the extension line of $\theta = \pi/2$. In (d), the velocity profiles in the center of the triplet circle boundary are shown, with the origin of the polar coordinate being set at the geometric center of the overlapping three circles. The geometric constraint was a circle having radius $R=$200, 250, and \SI{300}{\micro\meter} with $\Delta/R=$0.83, 1.44, and 1.83.}\label{fig4}
\end{figure*}

Furthermore, topological defects in active nematics can move spontaneously, with this motion often causing the annihilation of defects, resulting in the expansion of the aligned orientation field (FIG. \ref{fig1}(a)-(c)). Indeed, characteristic patterns of the velocity field appear around topological defects (FIG. \ref{fig1}(j) and (k)), and the superposition of those velocity field can dictate the overall collective motion. By taking our geometric patterning in spatially constrained cells, we next investigated the confined collective motion coupled with defect pairing. In a doublet circle boundary, two $+1/2$ defects with tails facing each other under all $\Delta/R$ conditions formed an inward flow from the boundary to the center (FIG. \ref{fig3}(a)). A similar collective motion from each circle’s center to the geometric center also occurred at $+1/2$ defects in the triplet circular boundary (FIG. \ref{fig3}(b)). Because $+1/2$ topological defects were located close to the center of each circular confinement facing each other, the velocity divergence was mostly negative throughout the confined space in both the doublet and triplet circle boundaries (FIG. \ref{fig3}(c)). The velocity divergence averaged over a radius of $R=\SI{200}{\micro\meter}$ from the center of the circle was negative regardless of the geometrical conditions $\Delta/R$, indicating that an inward contractile flow robustly heads to the center of the constrained region. Interestingly, as $\Delta/R$ increased, the magnitude of the velocity divergence decreased, with its geometric dependence independent of radius $R$. As a result, the contractile flow out of the $+1/2$ defect enclosed within the circle was directed outward from the center of the circle, which cancelled out the negative divergence. Therefore, as $\Delta$ increased, the effect of the outgoing contractile flow became larger than that of the incoming flow from another $+1/2$ defect in the neighboring circle.

The activity of the intracellular myosin molecular motor serves as the driving force for the motion of the individual cells. We examined whether perturbation of myosin activity affected the contractile flow of collective motion. Even in cells treated with 10 $\mu$M blebbistatin (a myosin inhibitor), the pairing of $+1/2$ defects was still built within the doublet circle boundary (FIG. S1). However, the negative velocity divergence has a smaller magnitude, with its dependence on $\Delta/R$ parameters largely lost. Maintaining sufficiently high contractile activity of myosin was essential for geometrically dependent contractile flow.

To further investigate the contractile flow stemming from geometrically controlled defect pairing, we utilized nucleus tracking velocimetry analysis to record the trajectory of individual cell motion. Specifically, low-toxicity SiR-DNA was employed to track the motion of the cell nucleus (FIG. \ref{fig4}(a), Movie S5 for doublet circle boundary, and Movie S6 for triplet circle boundary). In the doublet circle boundary, we established a polar coordinate system with the center of each circle serving as the origin. Additionally, we defined the angle $\theta$ such that the center of the other circle lied on the extension line of $\theta = \pi/2$. The radial and angular velocity components, $\bm{v_r}$ and $\bm{v_{\theta}}$, respectively, represented the motion of the cell nucleus in proximity to topological defects. We found that the radial velocity $\bm{v_r}$ exhibited a clear angular dependence within the doublet circle boundary, displaying faster outward motion from the center at an angle of $\theta =\pi/2$ and faster inward motion toward the center at $\theta = 3\pi/2$ (FIG. \ref{fig4}(b)). We commonly observed this spatial profile in $\bm{v_r}$ under various geometric conditions, including $\Delta/R$ and the constraint radius $R$. 

Furthermore, the angular velocity $\bm{v_{\theta}}$ also displayed angular dependence, with slower motion in the angular direction at $\theta =\pi/2$. Given that a $+1/2$ defect was situated near the origin, as depicted in FIG. \ref{fig2}(b) and oriented in the direction of $\theta =\pi/2$, this angular dependence of cell motion suggested that individual cells move in the direction toward $+1/2$ defects, which faced each other (FIG. \ref{fig3}). In the triplet circle boundary, similar angular dependencies of $\bm{v_r}$ and $\bm{v_{\theta}}$ are observed, indicating that the $+1/2$ topological defect promotes directional control of cell motion and induces contractile flow of collective motion toward the center of the confined spaces (FIG. \ref{fig4}(c)).

Not only were $+1/2$ defects present, but another $-1/2$ defect was also situated near the geometric center of the triplet circle boundary (FIG. \ref{fig2} and \ref{fig3}). By designating the geometric center of the triplet circle boundary as the origin, we analyzed the radial and angular velocity components, $\bm{v_r}$ and $\bm{v_{\theta}}$, respectively, of the cell nucleus in proximity to the $-1/2$ defect. We found a three-fold symmetry pattern in $\bm{v_r}$ at $\Delta/R$ values of 1.44 and 1.83, reflecting the contractile flow emanating from the three $+1/2$ defects locate within the three circles. In contrast, no clear geometric dependence was observed in $\bm{v_{\theta}}$ (FIG. \ref{fig4}(d)), consistent with the fact that nematic cell interactions alone do not result in net angular flow within closed symmetric boundaries.

\begin{figure}[tb]
\begin{center}
\includegraphics[scale=0.58,bb=0 0 452 572]{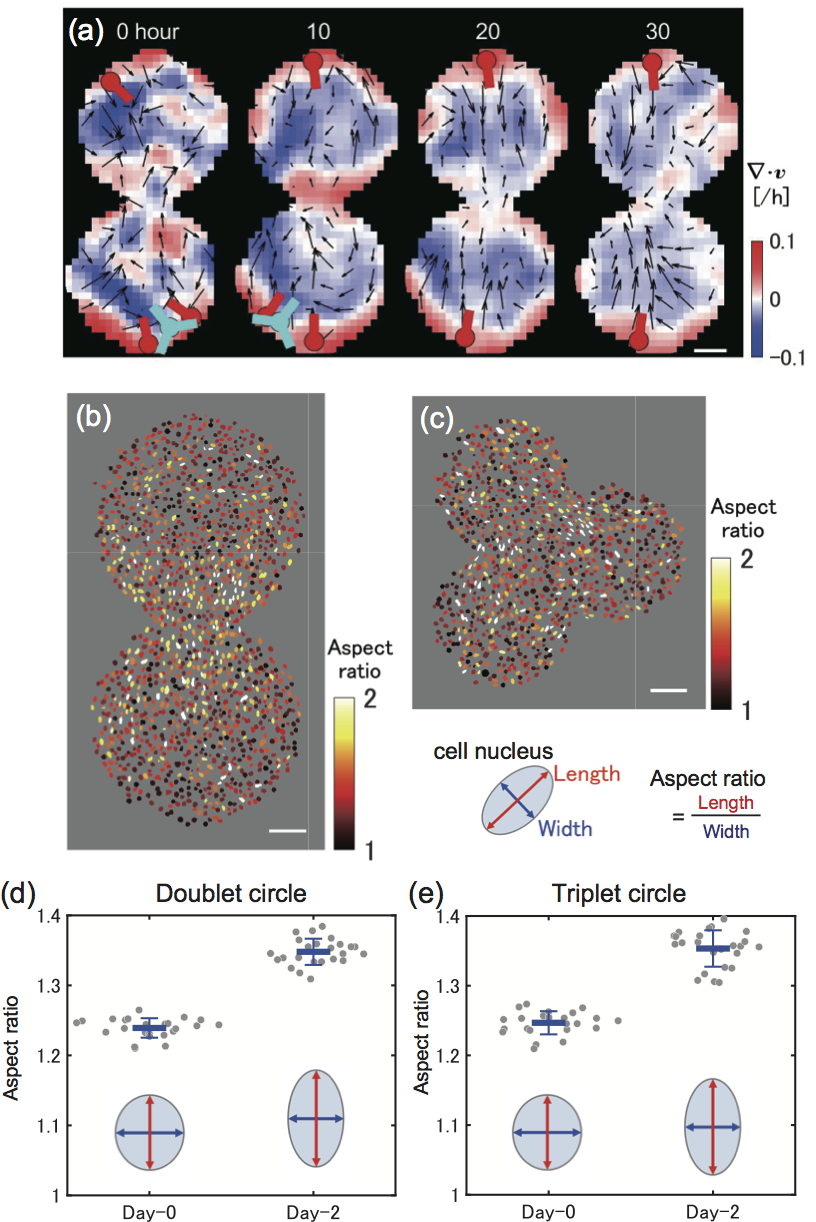}
\end{center}
\caption{\textbf{The deformation of cell nucleus under the pairing of topological defects.} Geometric parameters are $R$=\SI{300}{\micro\meter} and $\Delta/R$=1.83. Scale bars: \SI{100}{\micro\meter}. \textbf{(a)} Representative time evolution of velocity divergence of confined cells in the doublet circle boundary. Scale bar: \SI{100}{\micro\meter}. \textbf{(b and c)} The spatial map of the cell nucleus in the doublet circle boundary (b) and the triplet circle boundary (c). The representative fluorescent images taken at 48 hours are shown with color code for aspect ratio. The aspect ratio is defined as the ratio of major length (length) to minor length (width) of the cell nucleus. The large aspect ratio in white color indicates stretched nucleus shape. Cell nuclei with a high aspect ratio are found around the midpoint of the pair of $+1/2$ defects. \textbf{(d and e)} Temporal increase in aspect ratio of cell nuclei. The shape of cell nucleus at the start of the measurement $t=\SI{0}{\hour}$ (Day-0) is compared with that at $t = \SI{48}{\hour}$ (Day-2). The aspect ratio increases in both the doublet circle boundary (d) and the triplet circle boundary (e).}\label{fig5}
\end{figure}

Finally, to investigate its biological relevance, we examined the effect of contractile flow on cellular mechanics. In FIG. \ref{fig5}(a), the velocity divergence assumed negative values throughout the region over time, and the emergence of contractile flow became more pronounced between opposing $+1/2$ defects in the doublet circle boundary. As the cell population collectively moves in an aligned orientation, the contractile force exerted may alter the shape of the cellular structures, such as nucleus morphology. To answer this question, we analyzed the shape of the cell nucleus under geometric constraints at $\Delta/R=1.83$ and $R=\SI{300}{\micro\meter}$. By analyzing the aspect ratio (the ratio of the major length to the width) of the cell nucleus, we found that an elongated nucleus (i.e., high aspect ratio) was predominantly located near the midpoint of the pair of $+1/2$ defects, as well as the head part of $+1/2$ defects, in both the doublet (FIG. \ref{fig5}(b)) and triplet circle boundaries (FIG. \ref{fig5}(c)). Such nucleus deformation was not substantial at the initial stage (Day-0) but became more pronounced after approximately 48 hours (Day-2) (FIG. \ref{fig5}(d) and (e)). This result suggests that the contractile flow occurring in the region of the $+1/2$ defect pairing exerts mechanical stimulation on the cells, and the contractile forces accumulated over time deform the nucleus.

\subsection{Conclusion}

In this study, we investigated collective motion elicited by topological defects in a myogenic C2C12 cell population. The orientation field of the cell population generated topological defects on a two-dimensional plane, and the correlation length of the orientation field increased over time as $+1/2$ defects and $-1/2$ defects were annihilated. We examined the control of defect positioning by enclosing cells within a geometric boundary, in which two or three circles overlapped. The confined cells formed an orientation pattern with two $+1/2$ defects under a doublet circle boundary, with three $+1/2$ defects and one $-1/2$ defect under a triplet circle boundary. These orientation patterns were the most stable configurations with a net topological charge of $+1$, indicating that the symmetry of the geometrical boundaries determines the positioning and heading angles of the topological defects.

Utilizing geometrically designed boundary conditions, we aimed to clarify the self-organization of the collective motion induced by topological defects. The pairing of $+1/2$ defects with tails facing each other induced the contractile flow of collective motion characterized by negative velocity divergence. This contractile flow occurred broadly under all geometrical conditions of $\Delta$ and $R$. We previously demonstrated that under boundary geometries with overlapping circles, the cellular orientation shifted from parallel to the boundary in a perpendicular direction at $\Delta/R=\sqrt{2}$ in bacterial suspensions \cite{beppu, beppu2} and gliding microtubules \cite{beppu3}. The change in orientation of the cell population to become perpendicular to the boundary can be suppressed by the contractile flow between the pair of $+1/2$ defects, indicating that the absence of such a pattern transition in confined C2C12 cells resulted from the strong tendency of the cell population to orient along the axis facing the $+1/2$ defects. Investigating the relationship between the strength of the contractile flow and the shift in the transition point remains a future study for controlling the topology of active nematics.

Our findings also provided novel insights into myotube induction, as C2C12 cells fuse into myotubes following culture induction. Deformation of the cell nucleus was elicited by establishing $+1/2$ defect pairing, indicating that the cells under contractile flow were exposed to mechanical stress over time. A recent study indicated the deformation of the cell nucleus also plays a role in controlling cell differentiation, such as mechanochemical signaling \cite{ingber1997}. Stretching of the nucleus under contractile flow could offer new insights into the mechanism of force-induced morphogenesis. Not only for biological relevance, the geometry-assisted manipulation of active nematics can be extended to chemically powered particles \cite{bartolo2021prx, bartolo2021pnas} for a deeper understanding of defects pairing \cite{amin2020} and diverse structure of topological fields \cite{sokolov2019}.

\subsection{Materials and Methods}
\subsubsection{Cell culture}
C2C12 cells were cultured in MEM medium (20\% FBS, Glutamax, Sodium pyruvate, and antibiotics) at 37\(^\circ\)C and 5.0\% CO$_{2}$. C2C12 cell suspension was spread in 35 mm glass-bottom dishes at an initial density $1.5\times10^{5}$ cells/mL. The cell culture was incubated for \SI{1}{\hour} to allow cells to attach to the substrate surface. For individual cell tracking, the cell nucleus was visualized using 250 nM SiR-DNA (CY-SC007, Cytoskeleton, Inc.). Myosin inhibition was performed with 10 $\mu$M blebbistatin (Thermo Fischer). 

\subsubsection{Microfabrication}
The patterning of the geometry of the cell adhesion region was created using reverse microcontact printing of PDMS (Sylgard 184, Corning). First, a pattern with the geometry of the adhesion region was created with a PDMS device, and then plasma surface treatment was applied to adhere it to a glass-bottom dish. The MPC polymer (Lipidure, NOF Corporation) coating agent was poured into this area to form a non-adhesive area. Then, the PDMS device was peeled off to form an uncoated area where the cell could adhere to the substrate. The photomask for microfabrication was purchased from MITANI Micronics, and other conventional microfabrication protocols were described in previous studies \cite{beppu, beppu2, beppu3}.

\subsubsection{Microscopy}
Microscopic measurement was conducted using an epifluorescence microscope (IX73, Olympus) with a CMOS camera (Zyla 4.0, Andor) and an LED light source. The culture medium's temperature and CO$_{2}$ concentration were controlled at 37\(^\circ\)C and 5.0\%, respectively, by the stage-top incubator (STXG-IX3WX, TOKAI-Hit). The microscope stage was automatically position-controlled, whereas we programmed multi-position time-lapse measurements of microfabricated C2C12 populations using Metamorph software (Molecular Devices), as shown in a previous study \cite{shigeta}.

\subsubsection{Data analysis}
The microscopic images taken were analyzed using the MATLAB image processing toolbox (Mathworks). The particle image velocimetry plug-in used for the analysis of the velocity field of collective motion. The orientation field $\phi(\bm{r},t)$ was analyzed using a deep-learning-based image processing tool (nematic defect finder) \cite{chen} to identify the location and orientation of the half-integer topological defects. For the particle tracking velocimetry analysis of single cells, single cell nucleus was tracked using TrackMate7.0 \cite{ersho}, and their shapes were detected using Cellpose2.0 \cite{stringer}.

\subsection{Acknowledgement}
We thank K. Kawaguchi (RIKEN) and L. Yamauchi (RIKEN) for their technical advice. This work was supported by Grant-in-Aid for Scientific Research on Innovative Areas 16H00805 and 18H05427, Grant-in-Aid for Scientific Research (B) 20H01872, Grant-in-Aid for Challenging Research (Exploratory) 21K18605 from MEXT, and Sumitomo foundation for basic research.

\end{document}